\documentclass[a4paper]{spie}  

\usepackage{amsmath,amsfonts,amssymb}
\usepackage{graphicx}
\usepackage[colorlinks=true, allcolors=blue]{hyperref}
\usepackage{threeparttable} 
\usepackage{macro_ref}
\usepackage{geometry}
\usepackage{siunitx}
\title{Pushing high angular resolution and high contrast observations on the VLTI from Y to L band with the Asgard instrumental suite: integration status and plans} 

\author[a]{Marc-Antoine Martinod}
\author[b]{Michael J. Ireland}
\author[d]{Denis Defrère}
\author[c]{Stefan Kraus}
\author[a]{Frantz Martinache}
\author[e]{Peter G. Tuthill}
\author[g]{Emilie Bouzerand}
\author[e]{Julia Bryant}
\author[c]{Sorabh Chhabra}
\author[d]{Peter Chingaipe}
\author[b,p]{Benjamin Courtney-Barrer}
\author[e]{Fred Crous}
\author[a]{Nick Cvetojevic}
\author[e]{Daniel S. Dahl}
\author[f]{Colin Dandumont}
\author[h,i]{Steve Ertel}
\author[g]{Germain Garreau}
\author[g]{Adrian M. Glauser}
\author[k]{Xavier Haubois}
\author[j]{Lucas Labadie}
\author[a]{Stéphane Lagarde}
\author[c]{Daniel Lancaster}
\author[d]{Romain Laugier}
\author[a]{Roxanne Ligi}
\author[e]{Nathan Long}
\author[f]{Alexandra Mazzoli}
\author[d]{Thomas Mattheussen}
\author[m]{György Medgyesi}
\author[d]{Kwinten Missiaen}
\author[b]{Sébastien Morel}
\author[l]{Daniel J. Ahrer}
\author[e]{Barnaby Norris}
\author[c]{Jyotirmay Paul}
\author[a]{Romain Petrov}
\author[d]{Gert Raskin}
\author[a]{Sylvie Robbe-Dubois}
\author[e]{Gordon Robertson}
\author[d]{Muhammad Salman}
\author[n]{Ahmed Sanny}
\author[k]{Nicolas Schuhler}
\author[q]{Owain Snaith}
\author[e,o]{Adam K. Taras}

\affil[a]{Université Côte d'Azur, Observatoire de la Côte d'Azur, CNRS, Laboratoire Lagrange, France}
\affil[b]{Research School of Astronomy and Astrophysics, College of Science, Australian National University, Canberra 2611, Australia} 
\affil[c]{School of Physics and Astronomy, University of Exeter, Stocker Road, Exeter, EX4 4QL, United Kingdom}
\affil[d]{Institute of Astronomy, KU Leuven, Celestijnenlaan 200D, 3001 Leuven, Belgium}
\affil[e]{Sydney Institute for Astronomy, School of Physics, Physics Road, University of Sydney, NSW 2006, Australia}
\affil[f]{Space sciences, Technologies \& Astrophysics Research (STAR) Institute, University of Li\`ege, Li\`ege, Belgium}
\affil[g]{Institute for Particle Physics and Astrophysics, ETH Zurich, 8093 Zurich, Switzerland}
\affil[h]{Department of Astronomy and Steward Observatory, 933 North Cherry Ave, Tucson, AZ 89 85721, USA}
\affil[i]{Large Binocular Telescope Observatory, 933 North Cherry Ave, Tucson, AZ 85721, USA}
\affil[j]{I. Physikalisches Institut, Universit\"at zu K\"oln, Z\"ulpicher Str. 77, 50937 Cologne, Germany}
\affil[k]{European Organisation for Astronomical Research in the Southern Hemisphere, Casilla, 19001, Santiago 19, Chile}
\affil[l]{Max-Planck-Institut für Astronomie, Königstuhl 17, 69117 Heidelberg, Germany}
\affil[m]{Konkoly Observatory, Hungary}
\affil[n]{University of California, Los Angeles, United States}
\affil[o]{Leiden Observatory, Leiden University, PO Box 9513, 2300 RA Leiden, The Netherlands}
\affil[p]{Australian Astronomical Optics \& School of Mathematical and Physical Sciences, Macquarie University, Sydney 2109, Australia}
\affil[q]{Research Software Engineering, NUIT,  Newcastle University, Newcastle upon Tyne, NE1 7RU, UK}

\authorinfo{Further author information: Marc-Antoine Martinod: E-mail: marc-antoine.martinod@oca.eu}

\begin{document} 
\maketitle

\begin{abstract}
ESO’s VLTI has a history of record-breaking discoveries in astrophysics using high-angular resolution instrumentation. 
Its latest visitor instrument, the Asgard instrumental suite, is one key to further enhance the potential of the facility, particularly in the very near-infrared. 
It comprises four natively collaborating instruments: HEIMDALLR, a K-band fringe tracker, wavefront corrector and stellar interferometer in K band, with the same optics; Baldr, an H-band Zernike wavefront sensor; BIFROST, an Y-J-H-band photonic combiner whose main science case is studying the formation processes and properties of stellar and planetary systems; and NOTT, an L-band nulling interferometer for imaging young planetary systems.
Each of these instruments promise significant advances in their respective science goals that scale with their technical challenges and technology innovations.
The integration of Asgard is planned in three phases. 
The first one (integration, commissioning of HEIMDALLR and Baldr) is successfully done. 
In this paper, we show an overview of the current progress of the integration of Asgard, the first results of the on-sky commissioning of HEIMDALLR and the future steps and observing policies for Asgard to serve the broader community.
\end{abstract}


\section{Introduction}
\label{sec:intro}
The Very Large Telescope Interferometer (VLTI) and its second-generation instruments in delivering unique science has set European astronomy apart. 
The facility upgrades within the Gravity+ framework\cite{Eisenhauer2019} promise still further ground-breaking scientific discoveries. 
Leveraging these recent developments, the Asgard instrument suite will extend the scientific capabilities of the VLTI with a set of four instrument modules: BIFROST\cite{bifrost_kraus2026, bifrost_kraus2024} (Beam-combination Instrument for studying the Formation and fundamental paRameters Of Stars and planeTary systems) which is a Y/J/H-band combiner optimized for high spectral resolution, Baldr\cite{Courtney2024} which is an H-band injection optimizer for BIFROST, HEIMDALLR\cite{heimdallr_martinache2026_perfhdlr, ireland2018} (High-EfficIency Multiaxial Do-it ALL Recombiner) which is a high-sensitivity K-band fringe tracker, and NOTT\cite{defrere2018, hi5_defrere, defrere_l-band_2024} (Nulling Observations of dusT and planeTs) which is an L-band nuller optimized for high-contrast observations.
The integration of Asgard in VLTI is in three phases and this paper aims to highlight the progress and key results obtained so-far of instrumental and astrophysical interest.
Science cases and instrument capabilities have been extensively described in Martinod et al. (2023)\cite{Martinod2023JATIS}, Defrère et al. (2022)\cite{hi5_defrere} and Kraus et al. (2022)\cite{bifrost_kraus2024}.

\section{Observing with Asgard and Science goals}
\label{sec:science}
Asgard science cases are diverse and mostly revolve around stellar physics, multiple systems, and exoplanetary science\cite{bifrost_kraus2024, hi5_defrere}. 
Despite being an instrument visitor, non-members of the consortium could collaborate with the PIs of the instruments on the following themes:
\begin{itemize}
    \item Formation process of binary systems (HEIMDALLR and BIFROST)
    \item Mass accretion and ejection (BIFROST)
    \item Formation and evolution of exoplanetary systems and exoplanet atmospheres (NOTT)
    \item Protoplanetary and circumplanetary disks (BIFROST and NOTT)
    \item Exozodiacal dust (NOTT)
\end{itemize}

\section{Overall architecture and expected performance}
\begin{figure}[h]
    \centering
    \begin{tabular}{cc}
         \includegraphics[width=0.48\textwidth]{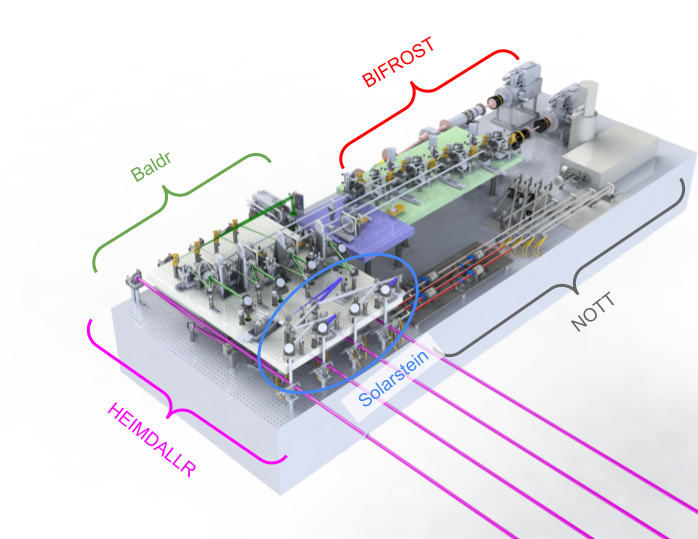} &
         \includegraphics[width=0.48\textwidth]{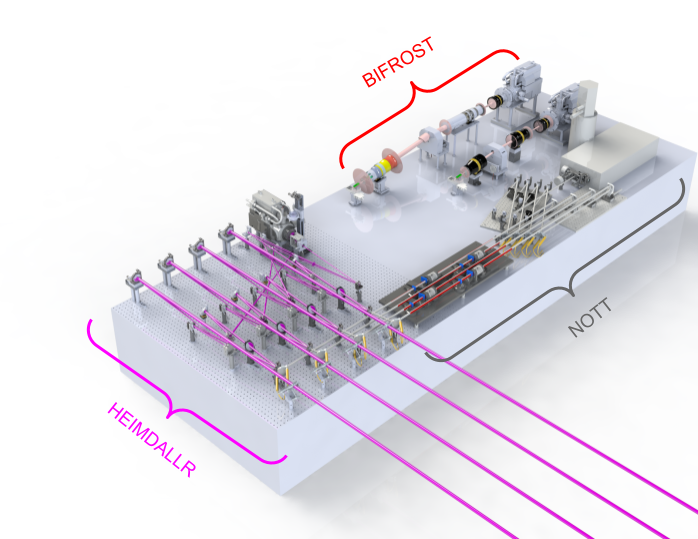} \\
    \end{tabular}
    \caption{Opto-mechanical design of Asgard. Left: The complete instrument suite, with the upper level shown. On this level, there are the optics of Baldr (Zernike wavefront sensor),  which shares a detector with HEIMDALLR. BIFROST (Y-H bands stellar interferometer) and Solarstein (calibration source unit generating calibration and alignment beams, utilizing injection mirrors to shift between the sky and internal source.) are also highlighted. Right: The lower level with the optics of HEIMDALLR (fringe tracker and stellar interferometer), BIFROST low- and high-resolution arms, and NOTT (L-band nulling interferometer) visible.}
    \label{fig:opto-mech}
\end{figure}

\begin{figure}[h]
    \centering
    \begin{tabular}{ll}
         \includegraphics[height=0.24\textwidth]{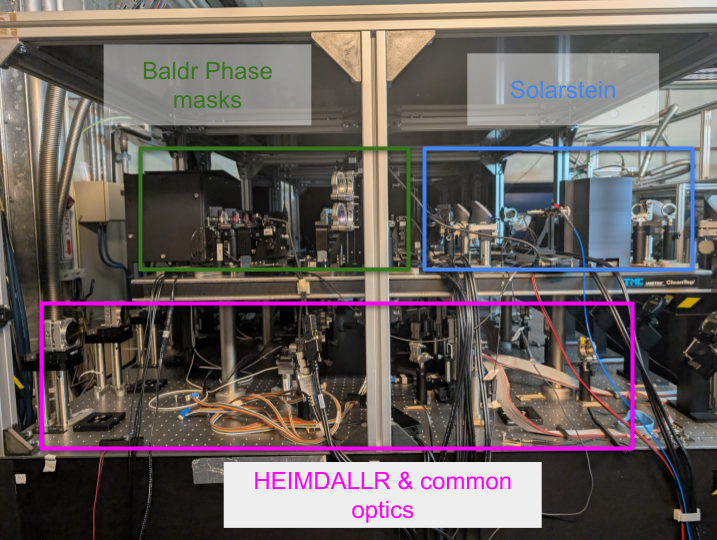} &
         \includegraphics[height=0.24\textwidth]{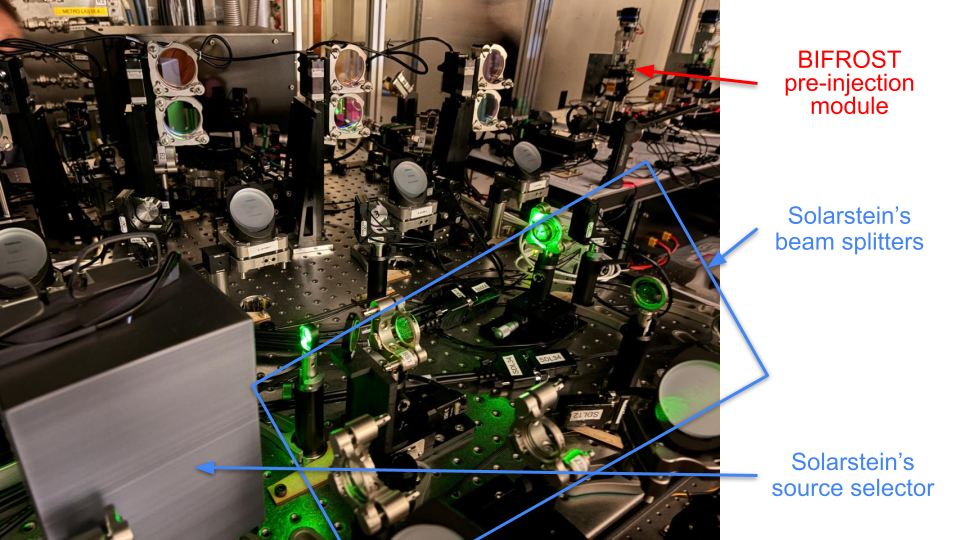} \\
         \includegraphics[height=0.24\textwidth]{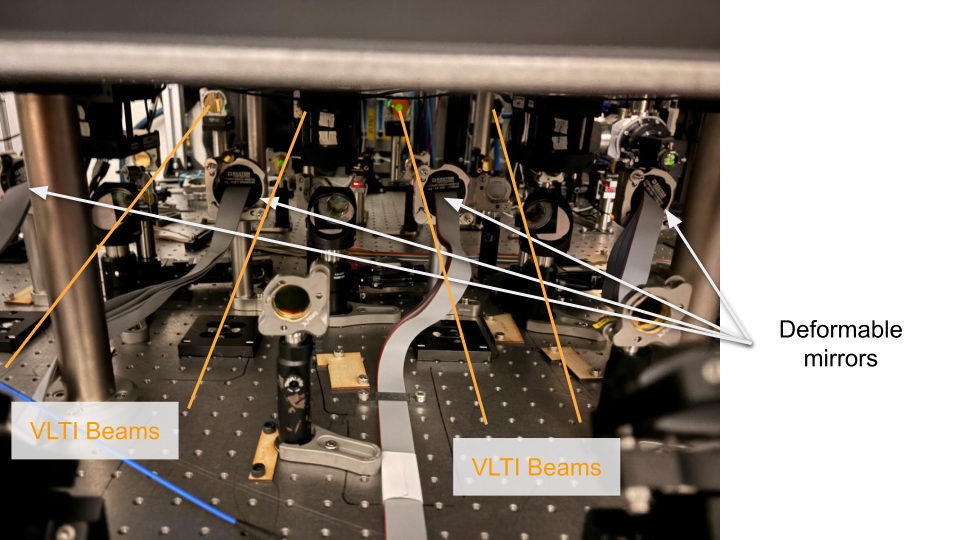} &
         \includegraphics[height=0.24\textwidth]{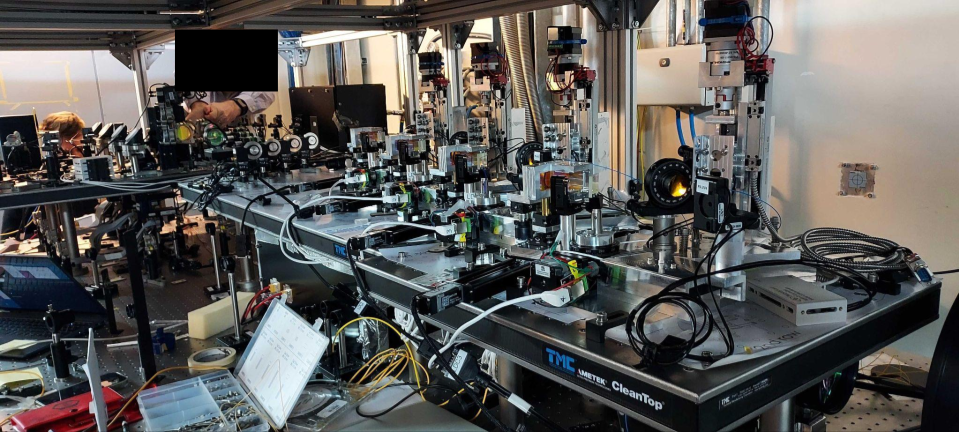}\\
    \end{tabular}
    \caption{Pictures of Baldr, Solarstein and HEIMDALLR's optics (top left), the upper level (top right), the point of view of the VLTI beams (bottom left) and BIFROST injection module (bottom right).}
    \label{fig:pictures}
\end{figure}

Asgard is located on the Visitor 2 table, formerly the AMBER table. 
The opto-mechanical design\cite{Martinod2023JATIS} of Asgard is shown in \autoref{fig:opto-mech}.
HEIMDALLR, Baldr and Solarstein are now integrated (Fig.~\ref{fig:pictures}).

\begin{figure}[h]
    \centering
    \includegraphics[width=0.9\textwidth]{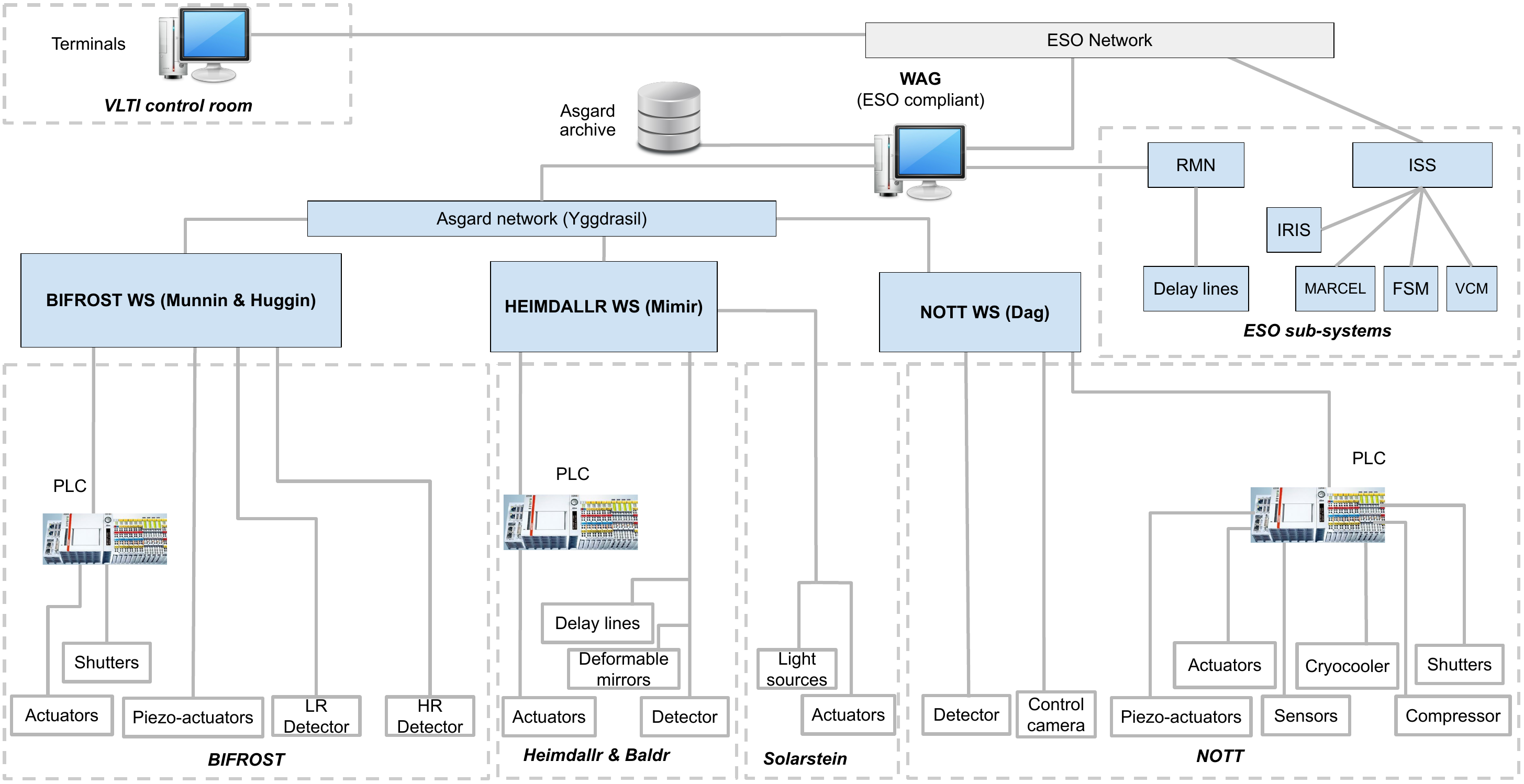}
    \caption{Control architecture of Asgard. WAG is an ESO-compliant machine that can communicate with ESO facility (in particular the RMN and the ISS) and control every Asgard subsystems through their respective workstations. Each instrument has one or two workstations to control its optics and cameras via programmable logic controllers (PLC). The users operate Asgard from terminals located in the VLTI control room that will connect to the WAG workstation.}
    \label{fig:asgard_architecture}
\end{figure}

The overall Asgard control architecture is given in \autoref{fig:asgard_architecture}.
The hardware architecture is designed so that all the modules can be operated from the terminals in the VLTI control room.
This centralisation allows the observer to run all the instruments simultaneously and to communicate with the VLTI systems. 

Asgard low-level IT hardware and software are non-ESO compliant; hence they must be kept separate. 
In addition, they need to communicate with each other; hence, they are connected to each other on a server-client system. 
Asgard modules need to interact with ESO systems such as the Reflective Memory Network (RMN), the delay lines and the calibration sources.
The solution is to connect all instruments’ workstations, via a local network to an ESO compliant machine, the instrument Workstation of AsGard, (WAG, formerly Hermod workstation) which is on ESO’s network. 
WAG will interact with ESO systems, the RMN, and transmit data to and from Asgard's modules that require such data. 
WAG respects ESO’s standards for hardware and software. 
From the ESO network point of view, Asgard is operated from a single computer represented by WAG.
Each instrument has one or two workstations called \emph{Module Control Units} (MCU):
\begin{itemize}
    \item The HEIMDALLR module control unit, named Mimir, reads out the image frames from HEIMDALLR \& Baldr's C-Red-1 camera, processes them in real-time, and sends out Optical Path Difference (OPD) and wavefront correction data. Mimir directly controls the Deformable Mirrors (DMs) and the internal delay lines within HEIMDALLR;
    \item The NOTT module control unit, named Dag, controls the camera and the opto-mechanical devices of NOTT;
    \item The first BIFROST module control unit, named Munnin, controls the C-Red-1 low-spectral-resolution camera and records its data. It also controls the on-axis beam opto-mechanical devices of BIFROST;
    \item The second BIFROST module control unit, named Huginn, controls the C-Red-1 high-spectral-resolution camera and records its data. It also controls the off-axis beam opto-mechanical devices of BIFROST.
\end{itemize}

Each instrument (except Solarstein) has its own data storage embedded in its respective MCU. 
In a future upgrade, the Asgard archive could also serve as the hub for feeding Asgard data into the ESO archive. 

The Asgard top-level control software on WAG bridges the gap between the Non-VLT-Software MCUs and the VLTI sub-systems. 
Providing a user interface common to VLT and VLTI instruments, it includes the following software components: an Observation Software, an Instrument Control Software and a Detector Control Software.

\section{Integration and development status}
\subsection{Integration status at VLTI}
\label{sec:integration}
The integration of Asgard followed a tier-based approach with three phases, each with an assembly and integration (AIV) period and a commissioning period.
The Phase 1 happened in July (AIV) and September (commissioning) 2025 and successfully ended February 2026.
It consisted of the HEIMDALLR and Baldr integration and commissioning as well as the setup of the Asgard infrastructure such as the WAG machine and the first templates.
Commissioning nights were dedicated to the tests of the fringe tracking capabilities with HEIMDALLR, adaptive optics control with Baldr and, visiblities and closure phase measurements.
Both modules have been tested on the AT and the UTs. 
On the UTs, the atmospheric dispersion correctors (ADCs) were successfully demonstrated.

The Phase 2 started in March 2026 with the integration of BIFROST.
This run highlighted the use of Baldr to improve BIFROST performance and validated the design changes Baldr had to undergo after Phase 1.
It will continue in August with the AIV of NOTT and in September with the commissioning of both modules.

The Phase 3 will start in the first semester of 2027 and see the commissioning of its high spectral resolution, off-axis modes, and photonic combiners, including a nulling combiner called Seidr\cite{dahl_seidr2026}.
The faint mode of observation of HEIMDALLR and Baldr along with BIFROST observing modes will be commissioned on the ATs and the UTs.

\subsection{Development status}
\subsubsection{BIFROST}
BIFROST\cite{bifrost_chhabra2026_design} will offer new and complementary benefits to the 2nd-generation VLTI instruments: higher angular resolution (due to shorter wavelengths), higher continuum flux for blue sources (as we observe closer to the peak wavelength of the photosphere), and a spectral resolution roughly 6 times greater (R=25,000) than what is currently available at the VLTI.
It will also give access to new line tracers, including Pa$\beta$ 1.282 $\mu$m, Pa$\gamma$ 1.094 $\mu$m, the He~I~1.083 $\mu$m accretion-tracing line\cite{Fischer2008}, and forbidden lines (e.g. [Fe II] 1.257 $\mu$m).
The instrument will include a fiber-fed all-in-one image plane combiner and PIC to best fit the science program.

The injection module and the on-axis optics\cite{bifrost_kraus2026, bifrost_jyotirmay2026_preinjection} have been installed on the Asgard bench (Fig.~\ref{fig:pictures}).
It encompasses the Longitudinal Dispersion Compensators which have been characterized \cite{bifrost_jyotirmay2026_ldc}, the shutters, the birefringence plates, the delay lines and the tip-tilt mirrors, with their electrical and firmware implementations\cite{bifrost_lancaster2026_electrical}.
The delay line actuators have been characterized\cite{bifrost_lancaster2026_DLM}. 
BIFROST has two spectrographs, namely a Low-resolution spectrograph\cite{bifrost_chhabra2026_lr_spectro} and a High-resolution arms with spectral resolution modes up to R=25,000. 
The low-resolution arm spectrograph is now integrated on the Asgard optical table and it contains a prism for R=50 with a Wollaston prism to measure separate polarization states, a VPH grating, both covering the Y/J-band or H-band.
For this one, we implemented an all-in-one beam combiner, although the system has initially also been designed to facilitate a photonic integrated circuit \cite{bifrost_lancaster2026_picsir}.
The LR arm measures OPD drift, dispersion, and fringe jumps to feed a feedback loop on the LDC and the BIFROST differential delay lines.
The BIFROST control software is adapted from the software framework used at CHARA/MIRC-X\cite{Anugu2020} to meet the needs of the new instrument\cite{bifrost_snaith2026_control_software}.
The control software must manage a large number of devices, including actuators and detectors, across the two arms of the instrument, as well as communicate with the gateway computer. 
Additional adaptations have been made to enable control via VLTI observing templates and interactions with the WAG interface and database.

\subsubsection{NOTT}
NOTT is the L-band nuller of Asgard. It follows a series of nulling interferometers that have been deployed over the past three decades on state-of-the-art facilities, both across single telescopes and as separate aperture interferometers\cite{Spalding2022}. Leveraging the experience gained with these nulling instruments, and thanks to the VLTI state-of-the-art infrastructure, NOTT has the potential to carry out several exoplanet programmes to study young Jupiter-like exoplanets at the most relevant angular separations. To achieve the contrast required to image young giant exoplanets, NOTT uses the so-called double-Bracewell architecture, with two stages of combiners. Several generations of photonic beam combiners made of Gallium Lanthanum Sulfide (GLS) were manufactured and tested \cite{nott_sanny2026_perf_chip}. The third iteration of the chip has been tested at cryogenic temperatures and in a vacuum. The tests show that the GLS photonic chip behaves as expected in such conditions, and its throughput is around 50\%\cite{nott_garreau2026_cryo_charac}. 
Nulling laboratory experiments show a broadband raw null\cite{nott_sanny2026_perf_chip} of $8.13 \pm 0.03 \times 10^{-3}$ and differential (by combination of its outputs) null depth of $1.14 \pm 0.01 \times 10^{-3}$\cite{nott_sanny2026_perf_chip, nott_mattheussen2026_labperf}. 
In 2025, the CO2 chambers \cite{Laugier2024}, used to compensate for the dispersion due to atmospheric CO$_2$, have been implemented. The first tests show they behave as expected, more quantitative ones are to come. Finally, the NOTT cryostat has been received in May 2026 and the project is currently performing the spectrograph integration \cite{Dandumont2022}. 
Temperature control has been validated, showing temperatures down to 30K and 100K, respectively for the detector and spectoragph at a pressure of 10$^{-8}$ mbar \cite{nott_chingaipe2026_camera}. 
Future tests will focus on the optical performance of the spectrograph at cryogenic temperatures.   

In parallel to the hardware development, a new data reduction pipeline has been developed \cite{Martinod2025} and a new data standard, called NIFITS\cite{nott_laugier2026_nifits}, has been designed. Staged-nullers, like double-Bracewell, encode information that cannot be properly stored in the OIFITS format, slowing its diffusion in the community. 
The new standard includes a high-fidelity parametric forward model of the instrument along with the data to be compatible with existing reduction and modeling tools\cite{nott_meilland2026_modelling} or custom-made by the user. 

\section{First on-sky commissioning of HEIMDALLR and Baldr}

\subsection{HEIMDALLR}
HEIMDALLR is a multi-axial beam combiner: the beams from the VLTI are placed next to each other in a compact two-dimensional non-redundant pattern and then brought to a common focus in a manner that inherits from sparse aperture masking interferometry. The fore-optics contained inside the so-called Narcissus box thoroughly described by Taras et al (2024)\cite{Taras2024}, located right before the light passes through the cold-stop of the instrument's CRED1 detector, split the K-beam into two sub-bands labeled K$_1$ ($\lambda = 2.05 \mu$m, with a band-pass of 0.2 $\mu$m) and K$_2$ ($\lambda$ = \SI{2.25}{\micro\metre}, with a band-pass of \SI{0.17}{\micro\metre}).

\begin{figure}
    \centering
    \includegraphics[width=\linewidth]{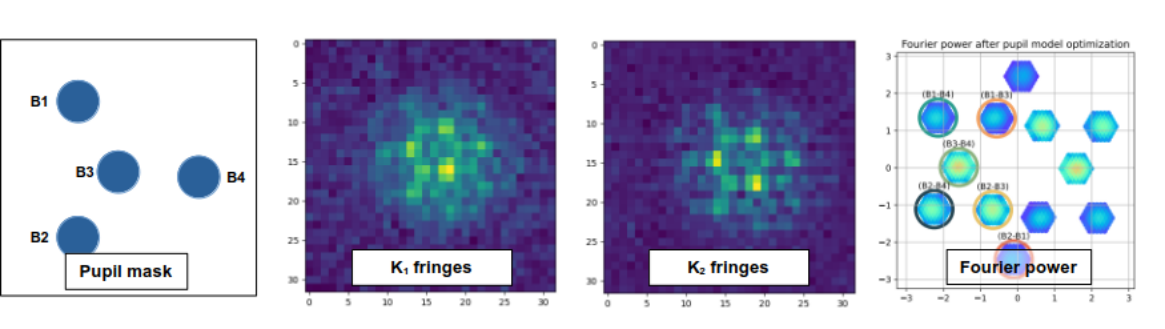}
    \caption{Fundamental wavefront sensing principle of HEIMDALLR. The beams from the VLTI (labeled B1, B2, B3 \& B4) are brought together in a 2D compact non-redundant array (first panel) and then to a common focus. The K-band beam is split into two K$_1$ and K$_2$ sub-bands (second and third panel) that are processed through 2D Fourier transform (fourth panel), giving access to astrophysically relevant interferometric quantities like V$^2$ and closure-phase along with inter- and intra-beam metrology information. The labels of the different splodges refer to the pair of beams that contribute to the complex visibility for said splodge.}
    \label{fig:hmd_principle}
\end{figure}

\autoref{fig:hmd_principle} illustrates the concept and features two simultaneous K$_1$ and K$_2$ interferograms acquired on-sky during early instrument commissioning with the VLTI auxilliary telescopes (ATs). To facilitate real-time processing and reduce the amount of data produced by a detector that can, given our current setup, run up to 3 kHz, the fringe tracking real-time and post-processing algorithms only use two 32x32 pixel regions of interest on the detector. 
In the standard 1 kHz mode, HEIMDALLR therefore processes $\sim$32 Mb of data per second, which on a busy night, can tally up to $\sim$.8 Tb. 
The two-dimensional information contained in these images gives access to multiple pieces of metrologic information due to the principle of the asymmetric pupil wavefront sensor \cite{2013PASP..125..422M} in addition to the interferometric observable quantities of astrophysical relevance. 
The real-time and/or post-processing of such data inherits from recipes developed in the context of sparse aperture masking interferometry. 
For our purpose here, it mostly relies on the 2D Fourier transform of both interferograms, leading to estimates of the complex visibility of the six possible baselines in the two bands, eventually getting six squared visibilities and up to four closure- or kernel-phases.

For fringe-tracking, the primary information comes from the phase of the splodges of information present in the Fourier transform of interferograms (see Fig.~\ref{fig:hmd_principle}) associated with the multiple baselines. 
If one notes $\gamma_1$ and $\gamma_2$ the complex visibilities measured in the respective band-passes, an estimate of the group delay is given by:

\begin{equation}
    \mathrm{GD} = \Lambda_{12} \times \arg{(\gamma_1 \times \gamma_2^{\star})} / 2\pi,
\end{equation}

\noindent
where $\Lambda_{12} = (0.5 (\lambda_1 + \lambda_2))^2/(\lambda_2-\lambda_1)$ is the unambiguous group-delay value capture range of the recombiner (in our case, $\Lambda_{12} = \SI{23}{\micro\metre}$, which is well matched to the coherence length of the K$_1$ band-pass). The group delay thus estimated for all six baselines is converted into a delay-line command (see Martinache, this conference), distributed between the VLTI delay-line at the 100-Hz correction rate and ASGARD's internal actuators. 

\begin{figure}
    \centering
    \includegraphics[width=.8\linewidth]{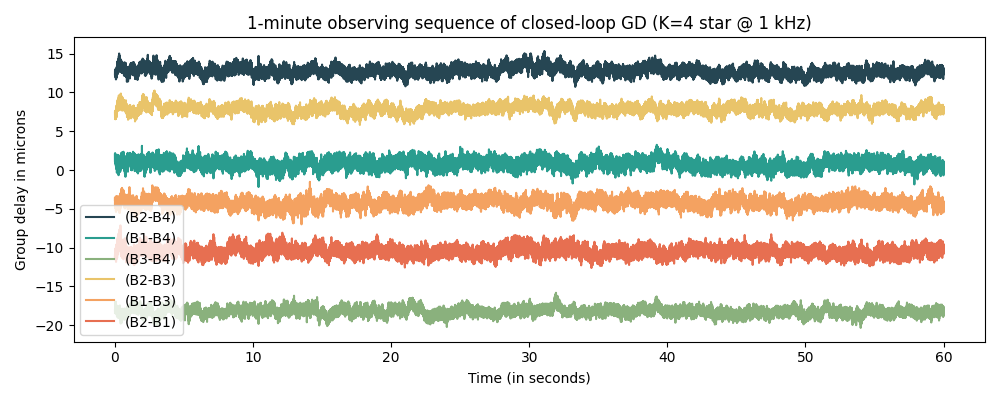}
    \caption{1-minute closed-loop time sequence (observing block) of group-delay error signals acquired by HEIMDALLR during commissioning, when observing a K=4.1 magnitude star (HD 90115) at 1 kHz frame rate.}
    \label{fig:gd_tracking}
\end{figure}

\autoref{fig:gd_tracking} shows an example of a 1-minute closed-loop time sequence\cite{heimdallr_martinache2026_perfhdlr} of group-delay error signals acquired by HEIMDALLR when observing a K=4.1 magnitude star (HD 90115) at a 1 kHz frame rate in the global correlated double sample (GCDS) mode of the CRED1 detector. The tracking performance is reasonably homogeneous across all the baselines, with residuals ranging from \SI{0.52}{\micro\metre} for the best baseline (labeled B3-B4 in Fig.~\ref{fig:gd_tracking}) to \SI{0.68}{\micro\metre} for the worst (B2-B1), with a global average of \SI{0.58}{\micro\metre} across multiple pointings of objects of similar brightness. 

The astute reader will observe that although indeed stable, the average value of the different group delays is not brought to zero by the control algorithm. While differential beam polarization plays a role (the instrument features four rotating LiNb0$_3$ plates in the common path to compensate for telescope-induced differential polarization), at the time these time sequences were acquired, HEIMDALLR was equipped with dichroic beam splitters designed to transmit light beyond the long end of the K-band to NOTT (and reflect the K-band towards the final optics of HEIMDALLR). 
Later investigation revealed that the dichroic on beam \#3 was faulty. 
A tiny ($\sim$5 nm) offset in the spectral response of the dichroic (likely due to delaminating of its coating) induced a large $\sim$\SI{20}{\micro\metre} chromatic differential delay between the two bands, making it impossible to simultaneously bring all the group delays to zero. 
The faulty dichroic has been replaced and the next ASGARD commissioning run that will follow the installation of NOTT will confirm the fringe-tracker's capability to drive the group delay to zero.

\begin{figure}
    \centering
    \includegraphics[width=.8\linewidth]{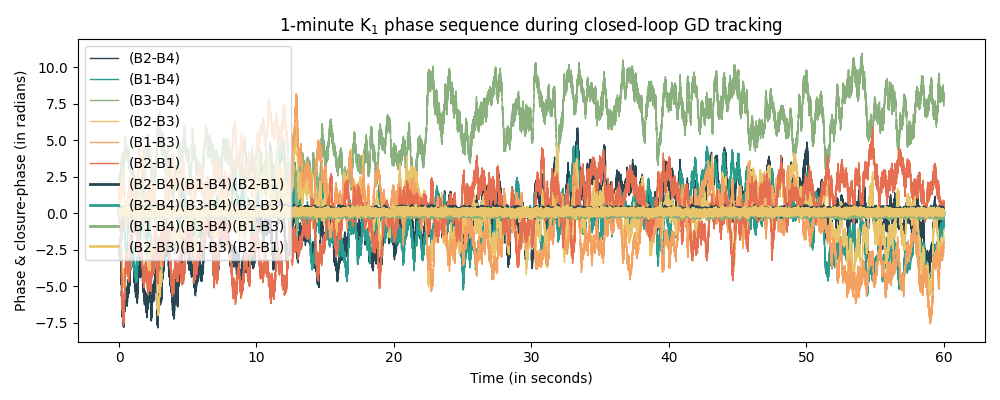}
    \caption{Evolution of the phase (in radians) measured during a 1-minute observing block in the K$_1$ band (same data sequence as for the group-delay plot of Fig.~\ref{fig:gd_tracking}) while group-delay tracking at 1 kHz frame rate. While the raw Fourier phase experiences significant fluctuations (average RMS across baselines $\sim$2.2 radians), it never ceases to amaze to see that simple linear well-chosen combinations of these noisy measurements lead to significantly more stable closure-phases!}
    \label{fig:phase_series}
\end{figure}

\begin{figure}
    \centering
    \includegraphics[width=0.7\linewidth]{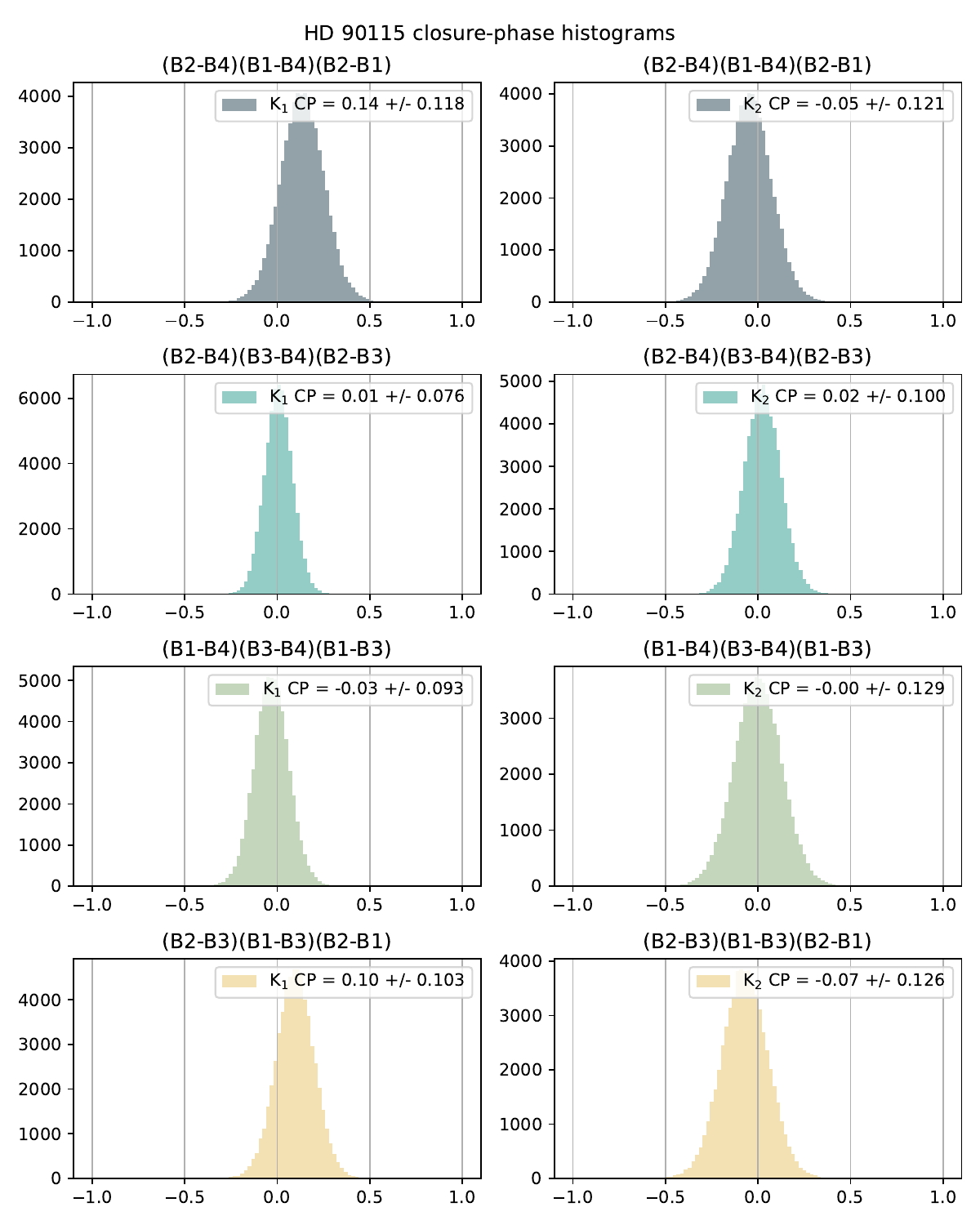}
    \caption{K$_1$ (left column) and K$_2$ (right column) histograms of all possible closure-phase acquired during the 1-minute observing block featured in Fig.~\ref{fig:gd_tracking} and Fig.~\ref{fig:phase_series}, based on 60,000 measurements. On a featureless calibration star like HD 90115, one expects to acquire closure-phases averaring to zero for all possible closure-triangles. For the most part, those closure-phase do indeed close, and follow normal statistics, with standard deviations that hover around 0.1 radians (about 5 degrees).
    Closure triangles involving beam \#1 (top and bottom row), more obviously in K$_1$, exhibit a definite small but non-zero instrumental closure-phase term, which we will be able to calibrate.}
    \label{fig:cp_histograms}
\end{figure}

Nevertheless, when taking advantage of this primary stabilization, the instrument can already acquire useful astrophysical observables\cite{heimdallr_martinache2026_perfhdlr}.
\autoref{fig:phase_series} shows how the still fluctuating phases acquired in the K$_1$ band when doing a group-delay tracking, with residuals ranging from $\sim$1.9 to 2.7 radians RMS, quickly leads to stable closure-phases. 
We can appreciate that in this bright star (K=4) use case, the closure-phase averages out very quickly, and a good estimate of the mean could be obtained from a significantly shorter observing block\cite{heimdallr_martinache2026_perfhdlr}.
\autoref{fig:cp_histograms} shows how these uncalibrated closure-phases (acquired on a featureless calibration star) are distributed. 
HEIMDALLR was specified (see Table 2 of Martinod et al, 2023\cite{Martinod2023JATIS}) to produce closure-phase estimates with a precision better than 1° ($\approx$0.017 radians).
The closure-phases clearly follow Gaussian statistics with a standard deviation of, in most cases, $\sigma \approx 0.1$ radians. 
An estimate of the mean (with uncertainty $\sigma_{CP} = \sigma/\sqrt{N}$, where $N$ is the number of frames used to evaluate the mean) could, under these observing conditions, be obtained over a just $\sim$ 50 milliseconds of observation. 

\begin{table}[]
    \centering
    \begin{tabular}{c|c|c|c|c}
         Target & CP$_{124}$ & CP$_{234}$ & CP$_{134}$ & CP$_{123}$\\
         \hline
         HD 90115 & 0.135 &  0.008 & -0.030 & 0.098 \\
         HD 90170 & 0.121 & -0.004 & -0.021 & 0.105 \\
         HD 90115 & 0.124 &  0.002 & -0.025 & 0.097 \\
         HD 90170 & 0.134 &  0.001 & -0.021 & 0.112 \\
         HD 90115 & 0.133 &  0.005 & -0.011 & 0.117 \\
         HD 90170 & 0.142 &  0.000 & -0.022 & 0.120 \\
         \hline
         RMS      & 0.007 & 0.004 & 0.006 & 0.009
    \end{tabular}
    \caption{Table of K$_1$ mean closure-phase (in radians) estimated after 1-minute observing blocks across multiple pointings of a pair of near equal brightness featureless (calibration stars), in the order of their acquisition. Each row lists the mean of the closure-phase 1-minute observing block for the four possible closure triangles. The final row (labeled RMS) estimates the deviation across all pointings.}
    \label{tab:cp_cal}
\end{table}

The uncalibrated closure-phases cannot, however, be used reliably on their own. 
We can indeed observe (see Fig.~\ref{fig:cp_histograms}) that two closure triangles, in particular in K$_1$ are affected by a systematic non-zero instrumental effect that offsets the location of the distributions. 
To calibrate the instrumental non-zero closure phase affecting its measurements, HEIMDALLR must, like most interferometers, be able to alternate between targets and featureless calibration stars of comparable brightness. 
The observing block that has been the focus of this section of the paper was part of a sequence of 6 pointings of the VLTI, alternating between two objects of near equal brightness (HD 90115 and HD 90170), a little over one degree apart in elevation. 
A summary of the closure phase extracted out of these 6 consecutive observing blocks (for K$_1$ only, but K$_2$ exhibits very similar properties) is provided in \autoref{tab:cp_cal}. 
The non-zero closure-phase diagnosed earlier proves to be very stable across multiple pointings. The usual closure-phase mutual subtraction calibration strategy would work to a remarkable degree. 
To estimate the error on the calibrated closure-phase, we can simply look at the standard deviation of the measurements listed in \autoref{tab:cp_cal}. 
The K$_1$ measurement with the largest deviation ($\sigma_{CP} = 0.009$ radians, that is $\sim$0.5°) is well within the 1° precision required by the primary observing program of HEIMDALLR. 
The K$_2$ closure-phase data that was simultaneously acquired is only very marginally better, so the same conclusion holds.

\begin{figure}
    \centering
    \includegraphics[width=0.8\linewidth]{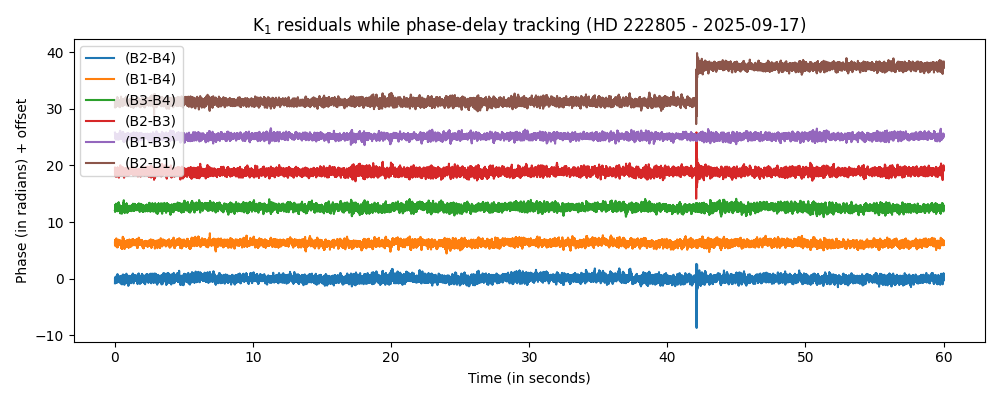}
    \includegraphics[width=0.8\linewidth]{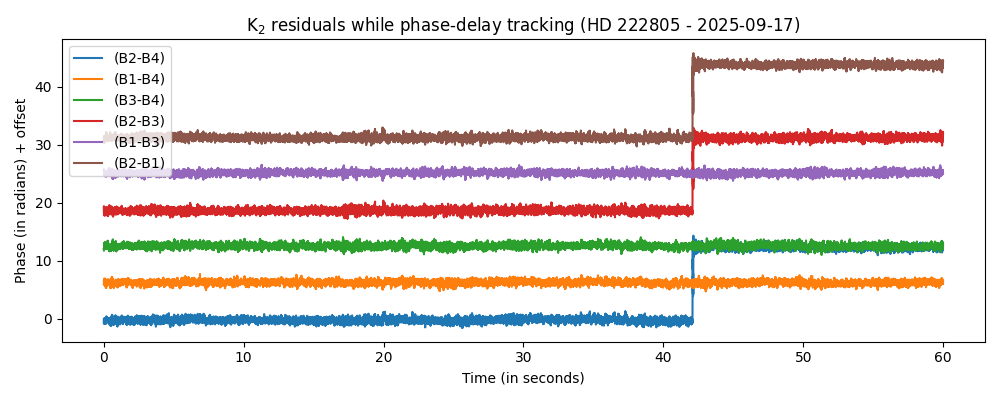}
    \caption{Example of phase-delay tracking residuals recorded by Heimdallr during a one-minute observing block of the target HD 222805 (K=4.2, observed on September 17, 2025). The top panel shows the phase residuals recorded by the K$_1$ channel. The bottom panel shows those recorded by the K$_2$ channel. The curves are here deliberately offset so as to better appreciate the relative performance on all baselines. The system experienced a fringe jump at the 41 second mark that is affecting all baselines involving Beam \#2.}
    \label{fig:pd_tracking}
\end{figure}

Upon successfully closing the loop on the group-delay, Heimdallr can also, when the observation conditions allow it, track and close a loop on the phase delay. The correction of the small phase-delay offsets is primarily handled by applying piston commands to the fast responding Asgard deformable mirrors. 
\autoref{fig:pd_tracking} shows one example of such a use case and plots for the two K$_1$ and K$_2$ channels, and the phase residuals recorded by all six baselines, during an observation of the K=4.2 magnitude star HD 222805, on September 17, 2025, with the ATs. For this particular early demonstration of phase-delay control, achieved during our first commissioning run, the phase RMS error is averaging $\sigma_{PD} = \SI{122}{\nano\metre}$. We're still not quite meeting the $\SI{50}{\nano\metre}$ target mark when observing with the ATs, but improvements in the control loop algorithm will get us there in time for the upcoming commissioning runs of BIFROST and NOTT.

\subsection{BALDR}
Baldr is the J- and H-band wavefront sensor of Asgard, providing wavefront control and injection stabilization for the wider instrument suite. Overview's of the instrument can be found in \cite{Taras2024, Courtney2024}. Following the first commissioning runs, the Baldr observing modes were revised to improve operational robustness. In particular, the original faint mode ZWFS concept was found to be strongly constrained by alignment tolerances. The faint mode has therefore been upgraded to a non-ZWFS configuration by adding an additional lens in the ZWFS focal plane, converting the ZWFS pupil plane image into a focal plane image. This provides a simpler and more robust faint target stabilization mode, while preserving a possible future path toward phase-diversity operation.

\begin{figure}[h]
    \centering
    \includegraphics[width=0.2\linewidth]{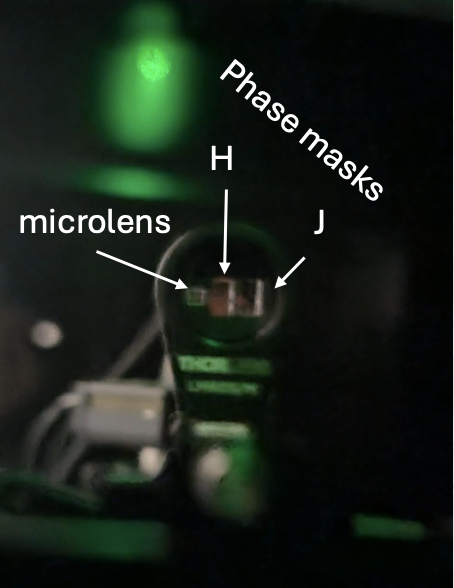}
    \includegraphics[width=0.7\linewidth]{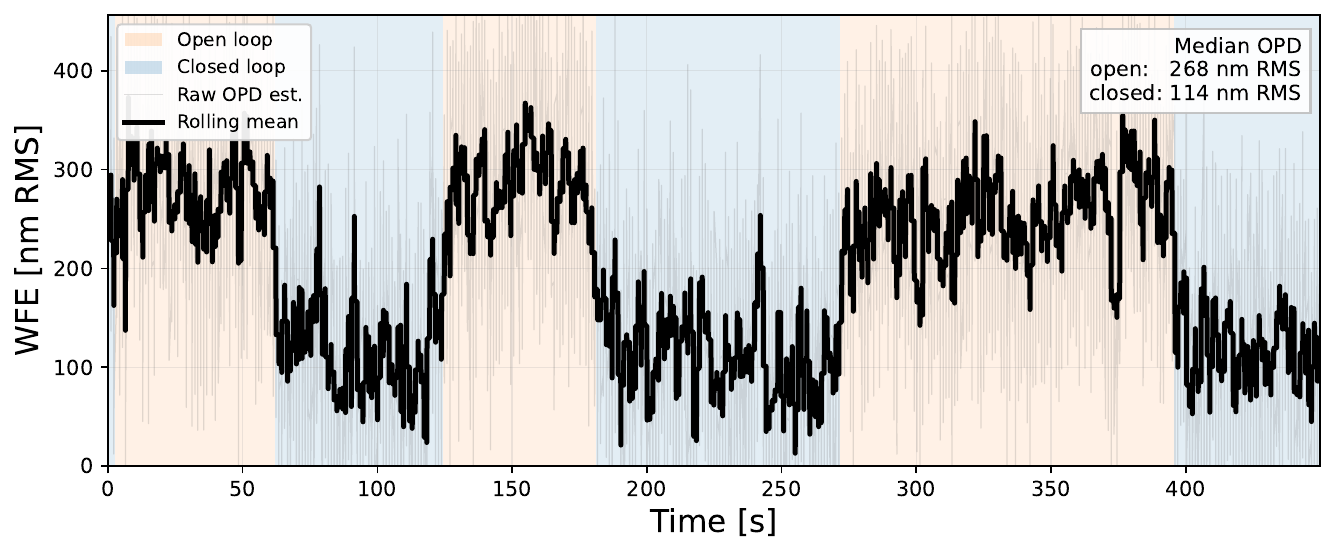}
    \caption{Left: upgraded Baldr optics in the focal plane containing new Zernike wavefront sensor phase masks for the bright mode of Baldr, optimized in the J- and H-bands respectively. Also a new microlens to upgrade the Baldr faint mode. Right: out-of-loop OPD estimate for Baldr while swapping between high-order servo (40 Zernike modes) on and off on the ATs. The WFE is estimated from the the intensity of light scattered outside the ZWFS pupil from light focused on the phase mask . HD 42165 (Hmag = 4.5), Baldr at 2kHz. Need to include pupils  }
    \label{fig:baldr_commissioning}
\end{figure}

The bright mode remains a ZWFS mode, but has also been upgraded. The pupil sampling was increased from 12 to 16 pixels per pupil diameter to improve high-order modal sensitivity. 
This comes at the cost of a modest reduction in limiting magnitude, shifting the transition between the bright and faint Baldr modes. 
The ZWFS phase masks were also redesigned: rather than using the previous lithographic SiO$_2$/SU8 process, the new masks are fabricated as etched silica optics, improving robustness and expected lifetime. 
Despite these upgrades, the bright mode still shows ongoing issues with repeatability and operational robustness. 
These limitations are likely linked to the tight internal alignment tolerances and intermediate plane vignetting in-addition coupled with the classic ZWFS challenge of variable optical gain / non-linearity – particularly around the pupil’s edge. This will remain a focus of continued commissioning campaigns. 
While these limitations prevent the bright mode from being considered fully operationally mature, the commissioning data nevertheless demonstrate that Baldr can achieve stable high-order loop closure under favourable alignment and observing conditions. 
\autoref{fig:baldr_commissioning} shows the upgraded Baldr focal plane optics and an example on-sky commissioning sequence for AT2. 
The sequence was obtained on HD42165 ($H=4.5$) with Baldr running at 2kHz, while alternating the high-order servo controlling 40 Zernike modes on and off. 
The out-of-loop OPD proxy, estimated from the light scattered outside the ZWFS pupil during the observation (and calibrated to OPD via a empirically fitted model on the internal source and the known DM influence function), decreases to near 114~nm RMS during the closed-loop intervals, in-line with requirements and predictions detailed in \cite{Taras2024,Courtney2024}. 
Further commissioning will continue improving repeatability and robustness of this mode.

\begin{figure}[h]
    \centering
    \includegraphics[width=0.7\linewidth]{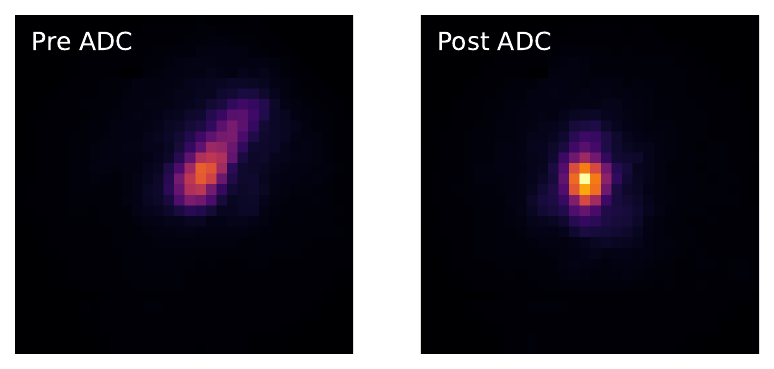}
    \caption{Atmospheric dispersion correct of a target at airmass 2.0 on the UTs. Images show the average of 1 second of image plane data at Baldr in J band. The ADC successfully reduces the dispersion at the shortest wavelengths used in the Asgard suite.}
    \label{fig:adc_commissioning}
\end{figure}

The faint image-plane mode has also been commissioned. Here we share preliminary results from the UTs, but the mode has also been tested on the ATs. 
As part of this, we also commissioned the atmospheric dispersion corrector (ADC) that is common between Baldr and Bifrost. \autoref{fig:adc_commissioning} shows the results before (left) and after(right) applying the ADC correction. 

\begin{figure}[h]
    \centering
    \includegraphics[width=0.95\linewidth]{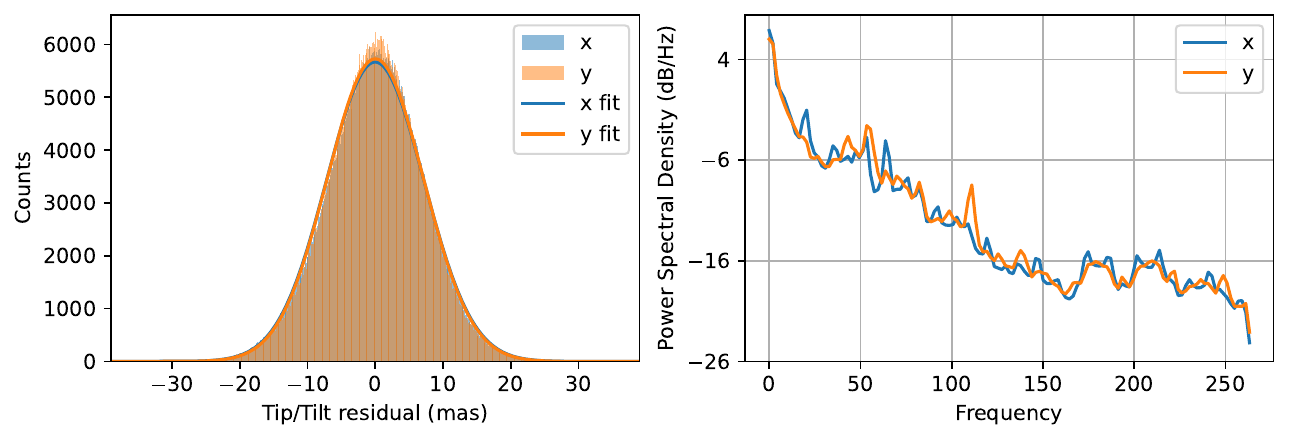}
    \caption{Tip-tilt residuals on a single beam with Baldr faint mode in closed loop. Histogram of residuals (left) shows mostly Gaussian behaviour in both axes with a standard deviation of around 7\,mas (0.55 pixels). Power spectral distribution (right) shows some differences between axes, however low frequencies dominate both axes. }
    \label{fig:baldr_faint_commissioning}
\end{figure}

We verified the Baldr faint mode on several targets, including SMSS J114447.77-430859.3\cite{onken2022discovery} ($m_J=12.8$mag) with the laser guide star mode in the new GPAO upgrade and Baldr's camera running at 500~Hz.
\autoref{fig:baldr_faint_commissioning} illustrates the performance of a single beam on this target. The results show a headline figure of around 7\,mas standard deviation per axis on this faint target. We observe structure in the power spectrum, indicating that further gain tuning and predictive control can improve the system.

\section{Conclusion}
The Asgard instrumental suite represents a major leap forward for the Very Large Telescope Interferometer (VLTI), opening new avenues for high-angular resolution and high-contrast astrophysics across the Y to L bands. 
The phased integration strategy has already demonstrated significant success, with Phase 1 successfully wrapping up in February 2026 after the integration and commissioning of HEIMDALLR and Baldr. 
On-sky commissioning results for HEIMDALLR have validated its compact, two-dimensional multi-axial beam combination design and confirmed its fringe-tracking robustness, achieving a global average group-delay tracking residual of 0.58 $\mu$m. 
HEIMDALLR has met its precision requirements on closure phase by delivering measurements with deviations well within $1^{\circ}$. 
Concurrently, the Baldr wavefront sensor has been successfully upgraded to optimize operational efficiency. 
Closed-loop sequences from Baldr demonstrate stable high-order loop corrections, reducing the out-of-loop wavefront error down to approximately 114~nm RMS under favorable conditions.  

Looking ahead, Phase 2 is actively underway following the initial integration of BIFROST in March 2026, which will be followed by the installation and commissioning of both BIFROST and NOTT later this year. 
BIFROST is poised to offer unparalleled spectral resolution (R=25,000), while its low-resolution arm is already integrated on the bench. 
Meanwhile, the NOTT L-band nulling interferometer continues to advance through cryogenic and laboratory validation, exhibiting promising self-calibrated null  depths on the order of $1.14 \times 10^{-3}$ using its GLS photonic chip. 
To facilitate widespread scientific adoption of these complex high-contrast observables, a new NIFITS data format is being designed. 
Finally, Phase 3 will start in the first semester of 2027 to finalize BIFROST’s high-spectral-resolution and off-axis modes alongside the faint-mode commissioning of the fringe tracker and wavefront sensor. 

\acknowledgments 
M-A.M \& F.M wish to acknowledge funding from the project PHOTONICS financed by the ANR program PEPR Origins (ANR-22-EXOR-0005).\\
M-A.M \& F.M wish to acknowledge funding from the Action Spécifique Haute Résolution Angulaire (ASHRA) of CNRS/INSU co-funded by CNES.\\
S. K. acknowledges support from the STFC Consolidated Grant (ST/V000721/1) and the STFC Small Award (ST/Y002695/1).\\
D.D. acknowledges support from the European Research Council (ERC) under the European Union's Horizon 2020 research and innovation program (grant agreement No. CoG - 866070).\\
S.K., S.C., J.P., and O.S. acknowledge support from an ERC Consolidator Grant (``GAIA-BIFROST'', grant agreement No.\ 101003096).\\
A.T., D.D., P.T., J.B., F.C. and B.N. acknowledge support from Astralis - Australia's optical astronomy instrumentation Consortium - through the Australian Government’s National Collaborative Research Infrastructure Strategy (NCRIS) Program as well as an Australian Research Council (ARC) Linkage Infrastructure Funding (LIEF) grant LE220100126. 
This work used the ACT, SA and Sydney nodes of the NCRIS-enabled Australian National Fabrication Facility.\\
D.J.A. and O.S. acknowledge support from the STFC Consolidated Grant (ST/V000721/1).\\
S.E. is supported by the National Aeronautics and Space Administration through the Astrophysics Decadal Survey Precursor Science program (Grant No. 80NSSC23K1473).

\bibliography{report} 
\bibliographystyle{spiebib} 

\end{document}